\documentclass[12pt]{article}

\usepackage{amssymb}
\usepackage{amsmath}
\usepackage{amscd}
\usepackage{latexsym}
\usepackage{graphicx}

\usepackage{cite}

\newcommand{\be}{\begin{equation}}
\newcommand{\ee}{\end{equation}}
\newcommand{\Dlt}{\Delta}
\newcommand{\dlt}{\delta}
\newcommand{\prt}{\partial}

\newcommand{\bt}{\beta}
\newcommand{\vp}{\varphi}
\newcommand{\ep}{\varepsilon}
\newcommand{\al}{\alpha}
\newcommand{\ra}{\rightarrow}

\newcommand{\gm}{\gamma}
\newcommand{\om}{\omega}

\newcommand{\lbd}{\lambda}

\begin{document}

\begin{center}

{\Large{\bf Calculating critical temperature and critical exponents by
self-similar approximants} \\ [5mm]

V.I. Yukalov$^{1,2}$ and E.P. Yukalova$^{3}$ }  \\ [3mm]

{\it
$^1$Bogolubov Laboratory of Theoretical Physics, \\
Joint Institute for Nuclear Research, Dubna 141980, Russia \\ [2mm]

$^2$Instituto de Fisica de S\~ao Carlos, Universidade de S\~ao Paulo, \\
CP 369, S\~ao Carlos 13560-970, S\~ao Paulo, Brazil \\ [2mm]

$^3$Laboratory of Information Technologies, \\
Joint Institute for Nuclear Research, Dubna 141980, Russia } \\ [3mm]

{\bf E-mails}: {\it yukalov@theor.jinr.ru}, ~~ {\it yukalova@theor.jinr.ru}

\end{center}

\vskip 1cm

\begin{abstract}

Self-similar approximation theory allows for defining effective sums of asymptotic 
series. The method of self-similar factor approximants is applied for calculating 
the critical temperature and critical exponents of the $O(N)$-symmetric $\varphi^4$ 
field theory in three dimensions by summing asymptotic $\ep$ expansions. This method 
is shown to be essentially simpler than other summation techniques involving 
complicated numerical calculations, while enjoying comparable accuracy.    
 
\end{abstract}

\vskip 2mm

{\bf Keywords}: Self-similar factor approximants; asymptotic series; critical exponents

\vskip 3mm

\section{Introduction}

Asymptotic expansions are widely employed in physics and applied mathematics 
\cite{Giacaglia_1,Nayfeh_2}. The expansion parameters are usually not small, 
because of which one needs to involve some kind of effective summation of divergent 
series. For this purpose, one often uses the method of Pad\'{e} approximants 
\cite{Baker_3}, Borel summation \cite{Kleinert_4} and its variants, such as 
Pad\'{e}-Borel summation and Borel summation with conformal mapping. One also uses 
the methods of renormalization group, conformal bootstrap, Monte Carlo simulations, 
and other methods (see \cite{Dupuis_5}) requiring quite heavy numerical calculations. 

In this communication, we advocate another approach based on self-similar approximation 
theory \cite{Yukalov_6,Yukalov_7,Yukalov_8,Yukalov_9}. The idea of this theory is to
represent the transition from one approximation term to another as the motion of a 
dynamical system, with the approximation order playing the role of discrete time. Then 
the sequence of approximation terms is bijective to the dynamical-system trajectory and 
the limit of the sequence is bijective to the fixed point of the trajectory. In the 
vicinity of a fixed point, the evolution equation acquires the form of a self-similar 
relation, which justifies the name of the self-similar approximation theory. Mathematical 
details can be found in the review articles \cite{Yukalov_10,Yukalov_11}. Here we only 
give the main ideas of the general theory in Sec. 2. In Sec. 3, we formulate the method 
of self-similar factor approximants. We do not plunge into the foundations of the theory, 
but just give the scheme of the method application. We apply the method for calculating 
the critical temperature shift for the $O(N)$ -symmetric $\varphi^4$ field theory in 
three dimensions in Sec. 4 and the critical exponents of this model in Sec. 5. We show 
that the method of self-similar factor approximants provides the accuracy comparable with 
other elaborate techniques involving heavy numerical calculations, while being essentially 
simpler. We compare our results with the most accurate values obtained by other methods 
summarized in Refs. \cite{Dupuis_5,Pelissetto_19,Abhignan_20,Shalaby_21}. We compare our 
results with Monte Carlo simulations, conformal bootstrap, hypergeometric Meijer summation, 
Borel summation, Borel summation with conformal mapping, and the method of nonperturbative 
renormalization group. This comparison shows good agreement of the self-similar factor 
approximants with the results obtained by other methods. Sec. 6 concludes.

\section{Self-similar approximation theory}

Let the function of interest be presented as an asymptotic expansion over a parameter 
$x$ assumed to be small,
\be
\label{1}
 f_k(x) = \sum_{n=0}^k a_n x^n \qquad ( x \ra 0 ) \;  .
\ee
For concreteness, we keep in mind a real function of a real variable. Although the 
expansion in powers of $x$ requires that the latter be asymptotically small, in reality
this variable is finite, and sometimes even is rather large. It is the standard situation
that series (\ref{1}) diverges, as $k$ increases, for any finite value of $x$. Hence,
it is necessary to have in hands a method for ascribing to the divergent series an 
effective sum. This can be done by resorting to self-similar approximation theory that 
is based on the following ideas (see reviews \cite{Yukalov_10,Yukalov_11}).

First, the given series has to be transformed by implanting in it control parameters,
\be
\label{2}
 \{ f_k(x) \} \mapsto \{ F_k(x,u_k) \} \;  .
\ee
These parameters are converted into control functions $u_k = u_k(x)$ whose role is to
make convergent the series $\{F_k(x,u_k(x))\}$.

Then the transition between the terms $F_k$ is reorganized so that to represent a motion
of a dynamical system, where the approximation order plays the role of discrete time,
\be
\label{3}
 F_k \mapsto  F_{k+1} \mapsto \ldots \mapsto  F^* \; .
\ee

The implantation of control functions can be done in different ways. For example, by 
incorporating control parameters into the initial step of perturbation theory. Thus, 
if the considered problem is characterized by a Hamiltonian $H$, one defines the 
Hamiltonian
\be
\label{4}
 H_\ep = H_0(u) + \ep [\; H - H(u) \; ] \qquad ( \ep \ra 1 ) \; ,
\ee
with a dummy parameter $\varepsilon$. Then perturbation theory is accomplished with 
respect to this dummy parameter. 

Another way of implanting control parameters is by a change of the variable, with a 
control parameter entering this change, and the following re-expansion in powers of
a new variable. For instance, one can substitute a change of the variable $x=x(z,u)$
into series (\ref{1}) and then re-expand it in powers of $z$ obtaining another series,
\be
\label{5}
\sum_{n=0}^k a_n x^n  ~ \mapsto ~ \sum_{n=0}^k b_n(u) z^n \;  .
\ee

Also, it is possible to transform the given series so that the transformed series 
would contain control parameters,
\be
\label{6}
 \hat T[\;u\;] \; f_k(x) = F_k(x,u_k) \; .
\ee

The convertion of control parameters into control functions is based on the Cauchy 
criterion of convergence, according to which the series converges if and only if for 
any positive $\varepsilon$ there exists $k^* = k^*(\varepsilon)$, such that
\be
\label{7}
|\; F_{k+p}(x,u_{k+p}) - F_k(x,u_k) \; | < \ep
\ee
for all $k > k^*$ and $p > 0$. The difference
\be
\label{8}
C(F_{k+p},F_k) \equiv F_{k+p}(x,u_{k+p}) - F_k(x,u_k)
\ee
is termed the Cauchy difference. The control functions are to be the minimizers of the 
convergence cost functional:
\be
\label{9}
 \min_u \; \frac{1}{2} \sum_k C^2(F_{k+p},F_k) ~ \mapsto ~ u_k(x) \;  .
\ee
  
The sequence of terms $F_k(x,u_k(x)$ is transformed into the trajectory of a dynamical 
system in the following way. We impose the rheonomic constraint
\be
\label{10}
F_0(x,u_k(x) ) = f 
\ee
that defines the expansion function
\be
\label{11}
 x = x_k(f) \; .
\ee
Let us introduce the endomorphism
\be
\label{12}
  y_k(f) : \; \mathbb{Z}_+ \times \mathbb{R} ~ \mapsto ~ \mathbb{R} \; ,
\ee
where
\be
\label{13}
  y_k(f) \equiv F_k(x_k(f),u_k(x_k(f) ) ) \;  .
\ee
The family of these endomorphsims forms a dynamical system in discrete time 
\be
\label{14}
 \{  y_k(f) : \; \mathbb{Z}_+ \times \mathbb{R} ~ \mapsto ~ \mathbb{R} \} \; ,
\ee
with the initial condition
\be
\label{15}
 y_0(f) = f \;  .
\ee
By construction, the dynamical system trajectory $\{y_k(f)\}$ is bijective to the 
sequence of approximants $\{F_k(x,u_k(x))\}$. If the limit of the approximation 
sequence exists, it is bijective to the trajectory fixed point
\be
\label{16}
 y_k(y^*(f) ) = y^*(f) \;  .
\ee
Since in practical applications we deal with finite approximation orders $k$, we 
obtain the approximants 
\be
\label{17}
  F^*_k(x) = F_k(x,u_k(x) ) \; .
\ee
    
In the vicinity of a fixed point, the self-similar relation
\be
\label{18}
y_{k+p}(f) = y_k(y_p(f) )
\ee
holds \cite{Yukalov_10,Yukalov_11}. The family $\{y_k(f)\}$, with relation (\ref{18}),
composes an approximation cascade. The approximation cascade can be embedded into an
approximation flow,
\be
\label{19}
 \{  y_k(f) : \; \mathbb{Z}_+ \times \mathbb{R} ~ \mapsto ~ \mathbb{R} \} \subset
\{  y(t,f) : \; \mathbb{R}_+ \times \mathbb{R} ~ \mapsto ~ \mathbb{R} \}  ,
\ee
which implies that the flow, which is a dynamical system in continuous time, enjoys the
self-similar relation
\be
\label{20}
y(t+t',f) = y(t,y(t',f) )
\ee
and the continuous flow trajectory passes through all points of the cascade trajectory,
\be
\label{21}
y(t,f) = y_k(f) \qquad ( t = k )
\ee
starting at the same initial point
\be
\label{22}
 y(0,f) = f \; .
\ee
 
The self-simlar relation (\ref{20}) can be represented as the Lie equation 
\be
\label{23}
\frac{\prt}{\prt t}\; y(t,f) = v(y(t,f) )
\ee
whose integration gives the evolution integral
\be
\label{24}
 \int_{F_k}^{F_k^*} \frac{dy}{v_k(y)} = t_k \;  .
\ee
The entering here velocity can be defined by means of the Euler discretization 
\be
\label{25}
 v_k(y) = y_{k+1}(f) - y_k(f) \;  .
\ee
The upper limit of the evolution integral gives the self-similar approximant
\be
\label{26}
 f^*_k(x) = F^*_k(x,u_k(x) ) \;  .
\ee

The procedure stability is characterized by calculating the local map multipliers
\be
\label{27}
  \mu_k(f) \equiv \frac{\prt y_k(f) }{\prt f} 
\ee
showing that the procedure is stable provided that
\be
\label{28}
 \mu_k < 1 \; , \qquad \mu_k \equiv \sup_f |\; \mu_k(f) \; | \; .
\ee
Respectively, the found fixed point is stable, when the fixed-point multiplier
\be
\label{29}
 \mu_k^*(f) \equiv \frac{\prt y_k^*(f) }{\prt f}
\ee
satisfies the condition
\be
\label{30}
 \mu_k^* < 1 \; , \qquad \mu_k^* \equiv \sup_f |\; \mu_k^*(f) \; | \;  .
\ee
Instead of local multipliers, one can consider the local Laypunov exponents
\be
\label{31}
\lbd_k(f) = \frac{1}{k} \; \ln \; | \; \mu_k(f) \; | 
\ee
in terms of which the stability condition reads as
\be
\label{32}
  \lbd_k < 0 \; , \qquad \lbd_k \equiv \sup_f  \lbd_k(f)  \;  .
\ee 

A particular example of transformation (\ref{2}) is the fractal transformation
\be
\label{33}
 F_k(x,s) = x^s f_k(x) = \sum_{n=0}^k a_n x^{n+s} \; ,
\ee
with the scaling exponent $s$ playing the role of a control parameter, and satisfying
the scaling relation
\be
\label{34}
\frac{F_k(bx,s)}{f_k(bx)} = \frac{F_k(x,s)}{f_k(x)} \; b^s \;   .
\ee

The rheonomic contsraint (\ref{10}) becomes
\be
\label{35}
  F_0(x,s) = x^s = f \;
\ee
giving the expansion function
\be
\label{36}
 x(f) = f^{1/s} \;  .
\ee

Endomorphism (\ref{13}) acquires the form
\be
\label{37}
 y_k(f) =  \sum_{n=0}^k a_n f^{1+n/s} \;  .
\ee
The cascade velocity (\ref{25}) reduces to 
\be
\label{38}
 v_k(f) = y_k(f) - y_{k-1}(f) = a_k f^{1+n/s} \;  .
\ee
 
Finding the fixed point $y_k^*(f)$ and accomplishing the inverse fractal 
transformation yields the self-similar approximant
\be
\label{39}
 f_k^*(x) = x^{-s} y_k^*(x^s) \;  .
\ee

\section{Self-similar factor approximants}

Employing the ideas described above, a convenient type of approximants has been 
derived, called self-similar factor approximants 
\cite{Yukalov_10,Yukalov_11,Yukalov_22,Gluzman_23}. Let an asymptotic expansion 
for a real function be given:
\be
\label{40}
f_k(x) = f_0(x) \left( 1 + \sum_{n=1}^k a_n x^n \right)
\ee
where a real parameter $x$ is assumed to be asymptotically small. However, we need 
to find the value of the function at a finite value of the parameter or even for 
$x \ra \infty$. In what follows, we assume that the sought function is sign-defined. 
Without the loss of generality, it can be treated as positive (non-negative). This is 
because a negative function $f$ can always be replaced by a positive function $-f$.  

The self-similar factor approximants, obtained from series (\ref{40}), read as  
\be
\label{41}
 f_k^*(x) = f_0(x) \prod_{j=1}^{N_k} ( 1 + A_j x )^{n_j} \;  ,
\ee
where the number of factors is
\begin{eqnarray}
\label{42}
N_k = \left\{ \begin{array}{rl}
k/2 ,     ~ & ~ k = 2, 4, 6, \ldots \\
(k+1)/2 , ~ & ~ k = 3, 5, 7, \ldots \end{array}
\right.   .
\end{eqnarray}
The parameters $A_j$ and $n_j$ are uniquely determined by the accuracy-through-order 
procedure, by equating the like-order terms in the expansions at small $x$,
\be
\label{43}
f_k^*(x) \simeq f_k(x) \qquad ( x \ra 0 ) \;   .
\ee
This procedure gives the equations
\be
\label{44}
 \sum_{j=1}^{N_k} n_j A_j^n = D_n \qquad ( n = 1, 2, \ldots , k)  \; ,
\ee
in which
\be
\label{45}
 D_n\equiv \frac{(-1)^{n-1}}{(n-1)!} \; \lim_{x\ra 0} \;
\frac{d^n}{dx^n}\; \ln \left( 1 + \sum_{m=1}^n a_m x^m \right) \;  .
\ee

For even orders $k$, Eq. (\ref{44}) consists of $k$ equations uniquely defining the 
$k/2$ parameters $A_j$ and $k/2$ parameters $n_j$. For odd orders $k$, the system 
of $k$ equations (\ref{44}) contains $k+1$ unknowns, where one of the parameters $A_j$, 
say $A_1$, is arbitrary. It is possible to use the normalization $A_1=1$, which makes 
the system of equations (\ref{44}) self-consistent and all parameters uniquely defined 
\cite{Yukalov_10,Yukalov_15}. If the parameters lead to a complex-valued approximant, 
it is replaced by the nearest real-valued approximant. The final result is given by 
the average between the last two approximants $[f_k^*(x)+f_{k-1}^*(x)]/2$ and the error 
bar is defined as the half-difference between the last two different approximants 
$[f_k^*(x)-f_{k-1}^*(x)]/2$.

\section{Critical temperature} 

The interaction strength of weakly interacting particles can be characterized by the 
coupling parameter, also called gas parameter,
\be
\label{46}
 \gm \equiv \rho^{1/3} a_s \;  , 
\ee
where $\rho$ is average density and $a_s$, scattering length. Many-body systems are 
often modeled by $O(N)$-symmetric $\vp^4$ field theory in three dimensions. When 
diminishing temperature, these systems exhibit a phase transition with spontaneous 
symmetry breaking. 

An interesting problem that has attracted much attention is the dependence of the 
phase transition temperature on the interaction strength 
(see reviews \cite{Andersen_2004,Yukalov_2004,Yukalov_2017}). One usually considers 
the critical temperature shift
\be  
\label{47}
\frac{\Dlt T}{T_0} \equiv \frac{T_c(\gm) - T_0}{T_0}
\ee
describing the relative difference between the critical temperature $T_c(\gm)$, as 
a function of the gas parameter, and the critical temperature $T_0$ of noninteracting 
particles, when $\gamma = 0$. At asymptotically weak interactions, the critical 
temperature shift behaves as
\be
\label{48}
 \frac{\Dlt T}{T_0} \simeq c_1 \gm \qquad ( \gm \ra 0 )\;  .
\ee
The problem is to find the coefficient $c_1$.

This coeficient has been calculated \cite{Kastening_1,Kastening_2,Kastening_3} in the 
loop expansion representing $c_1$ as an asymptotic series
\be
\label{49}
c_1 = c_1(x) \simeq \sum_{n=1}^5 a_n x^n   \qquad ( x \ra 0 )
\ee
in powers of the variable
\be
\label{50}
 x = (N + 2) \; \frac{\lbd_{eff} }{\sqrt{\mu_{eff}} } \;  ,
\ee
where $\lambda_{eff}$ is effective interaction strength and $\mu_{eff}$ is effective
chemical potential. The difficulty arises at the phase transition temperature, where 
the chemical potential $\mu_{eff}$ tends to zero, hence the expansion variable $x$ 
tends to infinity. Thus one comes to the necessity of giving to the value $c_1(\infty)$ 
some meaning. 

The problem can be solved by using self-similar factor approximants 
\cite{Yukalov_2017,Yukalov_2017a}. Then series (\ref{49}) is transformed into the
expression
\be
\label{51}
 c_1^*(x) = a_1 x \prod_{i=1}^{N_k} ( 1 + A_i x)^{n_i} \;  ,
\ee
with the parameters $A_i$ and $n_i$ defined by the accuracy-through-order procedure. 
At large $x$, expression (\ref{51}) behaves as
\be
\label{52}
 c_1^*(x) \simeq B_k x^{\bt_k} \qquad ( x \ra \infty) \;  ,
\ee
with the amplitude and exponent
\be
\label{53}
B_k = a_1 \prod_{i=1}^{N_k} A_i^{n_i} \; , \qquad 
\bt_k = 1 + \sum_{i=1}^{N_k} n_i \;   .
\ee
The required finiteness of $c_1$ implies that $\beta_k$ equals zero, which yields
\be
\label{54}
 c_1^*(\infty) = B_k \qquad ( \bt_k = 0 ) \;  .
\ee
This procedure can be complemented by a correction accelerating convergence 
\cite{Yukalov_2017,Yukalov_2017a}.     
      
The described scheme has been applied to expansion (\ref{49}) for the number of 
components $N=0,1,2,3,4$. The coefficients $a_n$ of the loop expansion (\ref{49}) can
be found in \cite{Yukalov_2017,Kastening_3}. The results for $c_1$ are presented in 
Table 1 and compared with the available Monte Carlo simulations for $N = 1$ \cite{Sun_2003}, 
for $N = 2$ \cite{Kashurnikov_2001,Arnold_2001}, and for $N = 4$ \cite{Sun_2003}.

\section{Critical exponents}

Critical exponents can be represented in the form of $\varepsilon$-expansions in powers 
of $\ep=4-d$, where $d$ is space dimensionality. For $O(N)$-symmetric $\vp^4$ field 
theory in three dimensions, the five-loop expansions can be found in the book 
\cite{Kleinert_24}. The summation of the five-loop expansions by means of self-similar 
approximants was considered in \cite{Yukalov_13} for all $N$. It has been shown that 
for $N=-2$ and $N \ra \infty$ self-similar approximants reproduce the exact values of 
the exponents. The results are very accurate for large $N\gg 1$, with the errors 
decreasing as $1/N$ with increasing $N$.  

Here we demonstrate that the accuracy of self-similar factor approximants for the 
$O(N)$ -symmetric $\varphi^4$ theory in three dimensions can be essentially improved 
by using the seven-loop $\varepsilon$-expansions that are known for $N = 1$ \cite{Ryttov_25} 
and that have been derived, using the seven-loop coupling-parameter expansions 
\cite{Schnetz_26}, for $N = 0,2,3,4$ \cite{Shalaby_27}. 

For $N = 0$, the expansions are
\be
\label{55}
\nu^{-1} = 2 - 0.25\;\ep - 0.08594 \;\ep^2 + 0.11443 \;\ep^3 - 0.28751 \;\ep^4 + 
 0.95613 \;\ep^5 - 3.8558 \;\ep^6 + 17.784 \;\ep^7  \; ,
\ee
\be
\label{56}
\eta = 0.015625\;\ep^2 + 0.016602\; \ep^3 - 0.0083675\; \ep^4 + 0.026505\; \ep^5 - 
 0.09073\; \ep^6 + 0.37851\; \ep^7 \; ,
\ee
\be
\label{57}
\om = \ep - 0.65625\; \ep^2 + 1.8236\; \ep^3 - 6.2854\; \ep^4 + 26.873\; \ep^5 - 
 130.01\; \ep^6 + 692.1\; \ep^7 \; .
\ee
 
For $N=1$, the $\ep$ expansions read as
\be
\label{58}
\nu^{-1} = 2 - 0.333333\;\ep - 0.11728\; \ep^2 + 0.12453\; \ep^3 - 
0.30685\; \ep^4 +  0.95124\; \ep^5 - 3.5726\; \ep^6 + 15.287\; \ep^7  \; ,
\ee
\be
\label{59}
\eta = 0.018519\; \ep^2 + 0.01869\; \ep^3 - 0.0083288\; \ep^4 + 0.025656\; \ep^5 - 
 0.081273\; \ep^6 + 0.31475\; \ep^7 \; ,
\ee
\be
\label{60}
\om = \ep - 0.62963\; \ep^2 + 1.6182\; \ep^3 - 5.2351\; \ep^4 + 20.75\; \ep^5 - 
93.111\; \ep^6 + 458.74\; \ep^7 \; .
\ee

The $\varepsilon$ expansions for $N=2$ are
\be
\label{61}
\nu^{-1} =  2 - 0.4\; \ep - 0.14\;\ep^2 + 0.12244\; \ep^3 - 0.30473\; \ep^4 + 
0.87924\; \ep^5 -  3.103\; \ep^6 + 12.419\; \ep^7 \; ,
\ee
\be
\label{62}
\eta = 0.02\; \ep^2 + 0.019\; \ep^3 - 0.0078936\; \ep^4 + 0.023209\; \ep^5 - 
0.068627\; \ep^6 +  0.24861\; \ep^7 \; ,
\ee
\be
\label{63}
\omega = \ep - 0.6\; \ep^2 + 1.4372\; \ep^3 - 4.4203\; \ep^4 + 16.374\; \ep^5 
- 68.777\; \ep^6 + 316.48\; \ep^7 \; .
\ee 

The $\varepsilon$ expansions for $N=3$ read as
\be
\label{64}
\nu^{-1} =  2 - 0.45455\; \ep - 0.1559\; \ep^2 + 0.11507\; \ep^3 - 
0.2936\; \ep^4 + 0.78994\; \ep^5 - 2.6392\; \ep^6 + 9.9452\; \ep^7 \; ,
\ee
\be
\label{65}
\eta = 0.020661\; \ep^2 + 0.018399\; \ep^3 - 0.0074495\; \ep^4 + 
0.020383\; \ep^5 -  0.057024\; \ep^6 + 0.19422\; \ep^7 \; ,
\ee
\be
\label{66}
\om = \ep - 0.57025\; \ep^2 + 1.2829\;\ep^3 - 3.7811\;\ep^4 + 13.182\;\ep^5 - 
 52.204\;\ep^6 + 226.02\;\ep^7  \; .
\ee 

The $\varepsilon$ expansions for $N=4$ take the form
\be
\label{67}
\nu^{-1} =  2 - 0.5\; \ep - 0.16667\; \ep^2 + 0.10586\; \ep^3 - 0.27866\; \ep^4 + 
0.70217\; \ep^5 -  2.2337\; \ep^6 + 7.9701\; \ep^7 \; ,
\ee
\be
\label{68}
\eta = 0.020833\; \ep^2 + 0.017361\; \ep^3 - 0.0070852\; \ep^4 + 0.017631\; \ep^5 - 
 0.047363\; \ep^6 + 0.15219\; \ep^7 \; ,
\ee
\be
\label{69}
\omega = \ep - 0.54167\; \ep^2 + 1.1526\; \ep^3 - 3.2719\; \ep^4 + 
10.802\; \ep^5 - 40.567\;\ep^6 + 166.26\; \ep^7 \; .
\ee 
 
We calculate the corresponding self-similar factor approximants $f_k^*(\ep)$, as
is explained in Sec. 3, and set $\varepsilon = 1$. The results of the calculations 
for the exponents are shown in Table 2 for $N=0$, Table 3 for $N=1$, Table 4 for $N=2$,
Table 5 for $N=3$, and Table 6 for $N=4$. We compare the results obtained by means of 
Factor approximants (FA) with the results of other methods: Monte Carlo simulations (MC) 
\cite{Hasenbusch_28,Hasenbusch_29,Hasenbusch_30,Clisby_31,Clisby_32,Hasenbusch_33}, 
Conformal bootstrap (CB) \cite{Showk_34,Shimada_35,Kos_36,Echeverri_37,Simmons_38}, 
Hypergeometric Meijer summation (HGM) \cite{Shalaby_27}, Borel summation complimented
by additional conjectures on the behavior of coefficients (BAC) \cite{Kompaniets_39}, 
Borel summation with conformal mapping (BCM) \cite{Kompaniets_40}, and Nonperturbative 
renormalization group (NPRG) \cite{Dupuis_5,Hasselmann_41,Rose_42,Polsi_43}.      

Other exponents, $\alpha$, $\beta$, $\gamma$, and $\delta$, are calculated using the 
relations
\be
\label{70}
 \al = 2 - 3\nu \; , \qquad \bt = \frac{\nu}{2}\; (1 +\eta) \; \, \qquad
\gm = \nu ( 2 -\eta) \; , \qquad \dlt = \frac{5-\eta}{1+\eta} \;  .
\ee
Table 7 summarizes these results.

\section{Conclusion}
 
The method of self-similar factor approximants is a very simple and convenient tool 
for the summation of asymptotic series. The basis of the method is the consideration 
of the transfer from one approximation term to another as a motion of a dynamical 
system, with the approximation order playing the role of time. The motion in the 
vicinity of a fixed point is described by a self-similar relation. The fixed point 
of the evolution equation defines the effective limit of the sequence. The method
can be applied to a large class of problems, as can be inferred from the review 
articles \cite{Yukalov_10,Yukalov_11} and the recent papers \cite{Yukalov_12,Yukalov}. 

The method of self-similar factor approximants is used for the summation of 
$\varepsilon$ expansions for the $O(N)$- symmetric field theory in three dimensions. 
The series of seventh order in $\varepsilon$ are employed. The method is shown to 
produce accurate approximations, at the same time being very simple. The results are 
compatible with other known methods of summation.

\vskip 5mm

{\bf Author Contributions}

Both the authors, V.I. Yukalov and E.P. Yukalova, equally contributed to this work.

\vskip 2mm

This research did not receive any specific grant from funding agencies in the public, 
commercial, or not-for-profit sectors.

\vskip 2mm

Tables 2 to 7 are reprinted from Physics Letters A, vol. 425 (2022) 127899, V.I. Yukalov 
and E.P. Yukalova, Self-similar sequence transformation for critical exponents, 

{\parindent=0pt
https://doi.org/10.1016/j.physleta.2021.127899, with permission from Elsevier 0375-9601/2021 
(License number 5274960037803).}

\newpage

\begin{table}[ht]
\centering
\caption{Coefficient $c_1$ of the critical temperature shift for different numbers 
of the field components $N$, found by means of self-similar factor approximants, and 
compared with Monte Carlo simulations. }
\vskip 3mm
\label{Table 1}
\renewcommand{\arraystretch}{1.25}
\begin{tabular}{|c|c|c|c|}  \hline
$N$  &  $c_1$         &     $Monte\; Carlo$                     \\  \hline
0    &  0.77 $\pm$ 0.03 &                                         \\ \hline
1    &  1.06 $\pm$ 0.05 & 1.09 $\pm$ 0.09 (\cite{Sun_2003})          \\ \hline
2    &  1.29 $\pm$ 0.07 & 1.29 $\pm$ 0.05 (\cite{Kashurnikov_2001})  \\
     &                  & 1.32 $\pm$ 0.02 (\cite{Arnold_2001})      \\ \hline
3    &  1.46 $\pm$ 0.08 &                                         \\ \hline
4    &  1.60 $\pm$ 0.09 & 1.60 $\pm$ 0.10 (\cite{Sun_2003})      \\ \hline
\end{tabular}
\end{table}

\begin{table}[ht]
\centering
\caption{Critical exponents for $N=0$, found by different methods: Self-similar 
factor approximants (FA), Monte Carlo simulations (MC), Conformal bootstrap (CB),
Hypergeometric Meijer summation (HGM), Borel summation with additional conjectures 
on the behaviour of coefficients (BAC), Borel summation with conformal mapping (BCM),
and Nonperturbative renormalization group (NPRG).}
\vskip 3mm
\label{Table 2}
\renewcommand{\arraystretch}{1.25}
\begin{tabular}{|c|c|c|c|}  \hline
$Method$ &  $\nu$         &     $\eta$     &    $\om$   \\  \hline
FA       &  0.5877 (2)    &   0.0301 (2)   &  0.821 (15)   \\ \hline
MC       &  0.5875970 (4) &   0.031043 (3) &  0.899 (12)  \\ \hline
CB       &  0.5877 (12)   &   0.0282 (4)   &    $-$      \\ \hline
HGM      &  0.5877 (2)    &   0.0312 (7)   &  0.8484 (17)   \\ \hline
BAC      &  0.5874 (2)    &   0.0304 (2)   &  0.846 (15)   \\ \hline
BCM      &  0.5874 (3)    &   0.0310 (7)   &  0.841 (13)   \\ \hline
NPRG     &  0.5876 (2)    &   0.0312 (9)   &  0.901 (24) \\ \hline
\end{tabular}
\end{table}

\begin{table}[ht]
\centering
\caption{Critical exponents for $N=1$, found by different methods listed in Table 2.}
\vskip 3mm
\label{Table 3}
\renewcommand{\arraystretch}{1.25}
\begin{tabular}{|c|c|c|c|}  \hline
$Method$ &  $\nu$         &     $\eta$     &    $\om$   \\  \hline
FA       &  0.6300 (3)    &   0.0353 (3)   &  0.808 (9)   \\ \hline
MC       &  0.63002 (10)  &   0.03627 (10) &  0.832 (6)  \\ \hline
CB       &  0.62999 (5)   &   0.03631 (3)  &  0.830 (2)     \\ \hline
HGM      &  0.6298 (2)    &   0.0365 (7)   &  0.8231 (5)   \\ \hline
BAC      &  0.6296 (3)    &   0.0355 (3)   &  0.827 (13)   \\ \hline
BCM      &  0.6292 (5)    &   0.0362 (6)   &  0.820 (7)   \\ \hline
NPRG     &  0.63012 (16)  &   0.0361 (11)  &  0.832 (14) \\ \hline
\end{tabular}
\end{table}

\begin{table}[ht]
\centering
\caption{Critical exponents for $N=2$, found by different methods listed in Table 2.}
\vskip 3mm
\label{Table 4}
\renewcommand{\arraystretch}{1.25}
\begin{tabular}{|c|c|c|c|}  \hline
$Method$ &  $\nu$         &     $\eta$     &    $\om$   \\  \hline
FA       &  0.6710 (3)    &   0.0372 (4)   &  0.809 (11)   \\ \hline
MC       &  0.67169 (7)   &   0.03810 (8)  &  0.789 (4)  \\ \hline
CB       &  0.67175 (10)  &   0.0385 (6)   &  0.811 (10)     \\ \hline
HGM      &  0.6708 (4)    &   0.0381 (6)   &  0.789 (13)   \\ \hline
BAC      &  0.6706 (2)    &   0.0374 (3)   &  0.808 (7)   \\ \hline
BCM      &  0.6690 (10)   &   0.0380 (6)   &  0.804 (3)   \\ \hline
NPRG     &  0.6716 (6)    &   0.0380 (13)  &  0.791 (8) \\ \hline
\end{tabular}
\end{table}

\begin{table}[ht]
\centering
\caption{Critical exponents for $N=3$, found by different methods listed in Table 2.}
\vskip 3mm
\label{Table 5}
\renewcommand{\arraystretch}{1.25}
\begin{tabular}{|c|c|c|c|}  \hline
$Method$ &  $\nu$         &     $\eta$     &    $\om$   \\  \hline
FA       &  0.7099 (1)    &   0.0372 (4)   &  0.7919 (3)   \\ \hline
MC       &  0.7116 (10)   &   0.0378 (3)   &  0.773      \\ \hline
CB       &  0.7121 (28)   &   0.0386 (12)  &  0.791 (22)     \\ \hline
HGM      &  0.7091 (2)    &   0.0381 (6)   &  0.764 (18)   \\ \hline
BAC      &  0.70944 (2)   &   0.0373 (3)   &  0.794 (4)   \\ \hline
BCM      &  0.7059 (20)   &   0.0378 (5)   &  0.795 (7)   \\ \hline
NPRG     &  0.7114 (9)    &   0.0376 (13)  &  0.796 (11) \\ \hline
\end{tabular}
\end{table}

\begin{table}[ht]
\centering
\caption{Critical exponents for $N=4$, found by different methods listed in Table 2.}
\vskip 3mm
\label{Table 6}
\renewcommand{\arraystretch}{1.25}
\begin{tabular}{|c|c|c|c|}  \hline
$Method$ &  $\nu$         &     $\eta$     &    $\om$   \\  \hline
FA       &  0.7459 (2)    &   0.0361 (4)   &  0.7913 (8)   \\ \hline
MC       &  0.750 (2)     &   0.0360 (3)   &  0.765 (30)   \\ \hline
CB       &  0.751 (3)     &   0.0378 (32)  &  0.817 (30)     \\ \hline
HGM      &  0.7443 (3)    &   0.0367 (4)   &  0.7519 (13)   \\ \hline
BAC      &  0.7449 (4)    &   0.0363 (2)   &  0.7863 (9)   \\ \hline
BCM      &  0.7397 (35)   &   0.0366 (4)   &  0.794 (9)   \\ \hline
NPRG     &  0.7478 (9)    &   0.0360 (12)  &  0.761 (12) \\ \hline
\end{tabular}
\end{table}

\begin{table}[ht]
\centering
\caption{Critical exponents for the three-dimensional $O(N)$-symmetric
$\vp^4$ field theory, calculated by means of self-similar factor approximants}
\vskip 3mm
\label{Table 7}
\renewcommand{\arraystretch}{1.25}
\begin{tabular}{|c|c|c|c|c|c|c|c|}  \hline
$N$ & $\al$     &  $\bt$    &  $\gm$    &  $\dlt$   &  $\nu$   &  $\eta$  &  $\om$   \\  \hline
0   & 0.2369    &  0.3027 &  1.15771  &  4.8247  &  0.5877  &  0.0301  &  0.821    \\ \hline
1   & 0.1100     &  0.3261 &  1.23776  &  4.7954  &  0.6300  &  0.0353  &  0.808  \\ \hline
2   & $-$0.0130  &  0.3480 &  1.31704  &  4.7848  &  0.6710  &  0.0372  &  0.809  \\ \hline
3   & $-$0.1297 &  0.3682 &  1.39339  &  4.7848  &  0.7099  &  0.0372  &  0.792  \\ \hline
4   & $-$0.2377 &  0.3864 &  1.46487  &  4.7910  &  0.7459  &  0.0361  &  0.791  \\ \hline
\end{tabular}
\end{table}

\end{document}